\documentclass[aps,floatfix,showpacs,superscriptaddress,showkeys]{revtex4}

\usepackage{amssymb,amsmath,amsfonts,latexsym,graphicx,epsfig}
\usepackage{color} 
    
\usepackage{titlesec}
\titleformat{\section}{\large\bfseries}{\thesection}{1em}{}

\newcommand{\bea}{\begin{eqnarray}}
\newcommand{\ena}{\end{eqnarray}}
\newcommand{\nn}{\nonumber\\}

\newcommand{\be}{\begin{equation}}
\newcommand{\en}{\end{equation}}

\newcommand{\ed}{\end{document}}

\newcommand{\slp}{p\kern-5pt/}

\begin{document}

\title{Form factors of the $B-S$ transitions 
in the covariant quark model}

\author{Aidos Issadykov}
\affiliation{Bogoliubov Laboratory of Theoretical Physics, 
Joint Institute for Nuclear Research, 141980 Dubna, Russia}
\affiliation{Faculty of physics and technical sciences, 
L.N.Gumilyov Eurasian National University, 010008 Astana, 
Republic of Kazakhstan}

\author{Mikhail A. Ivanov}
\affiliation{Bogoliubov Laboratory of Theoretical Physics, 
Joint Institute for Nuclear Research, 141980 Dubna, Russia}

\author{Sayabek K. Sakhiyev} 
\affiliation{Faculty of physics and technical sciences, 
L.N.Gumilyov Eurasian National University, 010008 Astana, 
Republic of Kazakhstan}

\begin{abstract}

In the wake of exploring uncertainty in the full angular distribution
of the $B\to K\pi+\mu^+\mu^-$ decay caused by the presence
of the intermediate scalar $K^\ast_0$ meson, we perform the straightforward
calculation of the $B(B_s)\to S$ ($S$ is a scalar meson) transition form factors
in the full kinematical region within the covariant quark model.
We restrict ourselves to the scalar mesons below 1 GeV:
$a_0(980)$, $f_0(500)$, $f_0(980)$, and $K^\ast_0(800)$. 
As an application of the obtained results we calculate
the widths of the semileptonic and rare decays
$B(B_s)\to S\ell\bar\nu$, $B(B_s)\to S\ell\bar\ell$ and 
$B(B_s)\to S\nu\bar\nu$. We compare our results with those
obtained in other approaches.

\end{abstract}


\maketitle

\section{Introduction}

Recently, much attention has been paid
to the rare flavor-changing neutral current decay 
$B\to K^\ast (\to K\pi)\mu^+\mu^-$. One of the reasons for this was
the first measurement of form-factor-independent angular observables 
performed by the LHCb Collaboration \cite{Aaij:2013qta,Aaij:2013iag}. 
It has been claimed
that there is a 3.7$\sigma$ deviation from the Standard Model (SM)
prediction for one of the angular observables. Much effort has been
spent to explain this deviation by invoking the effects of new physics (NP) 
(for example, see  
Refs.~\cite{Hurth:2013ssa,Descotes-Genon:2013vna,Datta:2013kja,Bobeth:2012vn,Altmannshofer:2013foa,Altmannshofer:2014rta,Mandal:2014kma} and references 
therein).
The main emphasis of the above-mentioned papers was 
on the search for the physical observables that have low sensitivity 
to the form factors.    

In addition to the NP effects, the uncertainties related to the presence
of the intermediate scalar resonance $K^\ast_0$ decaying into
$K\pi$ have been intensively discussed in the literature 
\cite{Lu:2011jm,Doring:2013wka,Meissner:2013pba,
Becirevic:2012dp,Matias:2012qz,Blake:2012mb,Das:2014sra}. 
A detailed analysis of the $B\to K^\ast_J(\to K\pi)\mu^+\mu^-$ decay 
in the higher kaon resonance region was done in Ref.~\cite{Lu:2011jm}. 
In many papers, the Breit-Wigner form for the $K\pi$ mass spectra was used. 
However, this assumption cannot be justified for the broad scalar resonances 
like the $K^\ast_0(800)$ meson. The improvement of the description was done 
in Ref.~\cite{Doring:2013wka} by invoking the chiral perturbation 
theory for the $K\pi$ interaction. This issue was also generalized to 
$B_s \to K\pi \ell\bar\nu$ in  Ref.~\cite{Meissner:2013pba}.

As is well-known, short-distance physics is under control in the
description of the rare $B$ decays, whereas 
the effects of long-distance physics described by the hadronic form factors
lead to large uncertainties since they involve  nonperturbative QCD.
The calculation of the $B\to K^\ast$ transition form factors 
have been performed in many theoretical approaches and models.  
We must mention some of them: light-cone QCD sum rules \cite{Ali:1999mm}, 
QCD sum rules \cite{Colangelo:1995jv}, the lattice-constrained
dispersion quark model \cite{Melikhov:1997wp}, the simple dipole
parametrization \cite{Aliev:2001fc}, perturbative QCD at large recoil region
\cite{Chen:2002bq}, the relativistic quark model \cite{Ebert:2010dv}, and the
Dyson-Schwinger equations in QCD \cite{Ivanov:2007cw}.

The $B_s$ and $D_s$ to $K^\ast_0(1430)$ transition
form factors were calculated in Ref.~\cite{Yang:2005bv}
within an approach  based on QCD sum rules.
The form factors for the $B\to K^\ast_0(1430)$ transition have been evaluated 
in the light-front quark model \cite{Chen:2007na}. 
The form factors of rare $B \to K_0^\ast (1430) \ell^+ \ell^-$ decay were
calculated in Ref.~\cite{Aliev:2007rq} within three-point QCD sum rules.
The $B\to S$ transition form factors  have been investigated 
in the light-cone sum rules approach \cite{Wang:2008da}.
The transition form factors of $B(B_s)$-mesons decay into a scalar meson 
were studied in Ref.~\cite{Li:2008tk} within the perturbative QCD approach.
With these form factors, the decay width and branching ratios of 
the semileptonic $B\to S\ell\bar\nu$ and rare $B\to S\ell^+\ell^-$ decays
have been calculated.
The rare semileptonic decays $B_s\to [f_0(980),K^\ast_0(1430)]\ell^+\ell^-$
and  $B_s\to [f_0(980),K^\ast_0(1430)]\nu\bar\nu$ were investigated
in Ref.~\cite{Ghahramany:2009zz} in the framework of the three-point 
QCD sum rules.
The $B_s\to f_0(980)$ transition form factors were computed 
in Ref.~\cite{Colangelo:2010bg} by using  light-cone
QCD sum rules at leading order in the strong coupling constant 
and an estimate of next-to-leading-order corrections. 
A QCD light-cone sum rule was also used to evaluate
the $B_s\to S$ form factors and
$B_s\to S \ell\bar{\nu}_\ell, \ell  \bar{\ell}\,\,(\ell=e,\mu,\tau) $ 
branching ratios in  Ref.~\cite{Sun:2010nv}.
The twist-3 light-cone distribution amplitudes (LCDAs) of 
the scalar mesons were investigated in  Ref.~\cite{Han:2013zg} 
within the QCD sum rules. As an application of those twist-3 LCDAs,
the $B \to S$ transition form factors were studied
by introducing proper chiral currents into the correlator.

Recently, the $B-S$ form factors for two scalar nonet mesons below and above 
1~GeV were calculated in Ref.~\cite{Wang:2014vra} by taking into account 
the perturbative  ${\mathcal{O}}(\alpha_s)$ corrections to the twist-2 terms 
using the light-cone   QCD sum rules. They were used in
Ref.~\cite{Wang:2014upa} to study the semileptonic $B\to S\ell\bar\nu_\ell$
and rare $B\to S\ell^+\ell^-, S\bar\nu\nu$ decays.

In the wake of exploring uncertainty in the full angular distribution
of the $B\to K^\ast (\to K\pi)\mu^+\mu^-$ decay caused by the presence
of the intermediate scalar $K^\ast_0$ meson, we perform the straightforward
calculation of the $B(B_s)\to S$ ($S$ is a scalar meson) transition form factors
in the full kinematical region within the covariant quark model.
We restrict ourselves to the scalar mesons below 1 GeV:
$a_0(980)$, $f_0(500)$, $f_0(980)$, and $K^\ast_0(800)$  \cite{Agashe:2014kda}.
Actually, the internal structure of these mesons is not yet well established
(see Refs.~\cite{Amsler:2013wea,Amsler:2004ps} for a review).
We will use the simple $\bar qq$ interpretation of 
the low-lying scalar mesons in our calculation. 
The calculated form factors are used to evaluate the branching fractions
of the decay $B(B_s)\to S \ell\bar\ell $, where $\ell = e,\mu,\tau$.
We compare our results with those obtained in other approaches. 

The paper is organized in the following manner. In Sec.~II we give the necessary
theoretical framework which includes the effective Hamiltonian,
its matrix element between the initial and final states, the definition
of the hadronic form factors and the helicity amplitudes.
In Sec.~III we briefly discuss our covariant quark model and calculate
the form factors of the transitions $B\to S \ell\nu_\ell$ and 
$B\to S\ell^+\ell^-$.  Finally, we present our numerical results
for the differential decay distributions and branching ratios.
We compare our findings with the results of other approaches.

\section{Effective Hamiltonian and form factors}

We start with the on-shell decays $B_d\to (K,K^\ast_0,K^\ast)\ell^+\ell^-$
which can be described by using the effective Hamiltonian for the $b\to s$
transition  \cite{Buras:1994dj,Buchalla:1995vs}. 
The effective Hamiltonian leads to the free-quark $b\to s l^+l^-$
 decay amplitude:
\bea
M(b\to s\ell^+\ell^-) & = &
\frac{G_F}{\sqrt{2}}\frac{\alpha\lambda_t}{2\pi}\,
\left\{
 C_9^{\rm eff}\,
       \left(\bar{s}O^\mu b \right) \,\left(  \bar\ell\gamma_\mu \ell\right)
 + C_{10}\left(\bar{s}O^\mu b \right) \,
         \left(\bar\ell\gamma_\mu\gamma_5 \ell\right)
\right.
\nn
&-& \left. \frac{2\hat m_b}{q^2}\,C_7^{\rm eff}\,
\left( \bar{s}\,i\sigma^{\mu \nu}\,(1+\gamma^5)\, q^\nu \,b \right)\,
                        \left( \bar\ell\gamma_\mu\ell \right) \right\} 
\label{eq:free}
\ena
where $O^\mu = \gamma^\mu(1-\gamma^5)$
is the weak Dirac matrix, $\lambda_t=|V^\dagger_{ts}V_{tb}|$
is the product of the Cabibbo-Kobayashi-Maskawa elements, 
and $C_7^{\rm eff}= C_7 -C_5/3 -C_6$.
The Wilson coefficient $ C_9^{\rm eff}$ effectively takes
into account (i) the contributions from the four-quark
operators and (ii) the nonperturbative
effects coming from the $c\bar c$-resonance contributions
which are as usual parametrized by a Breit-Wigner ansatz \cite{Ali:1991is}:
\bea
C_9^{\rm eff} & = & C_9 +
C_0 \left\{
h(\tilde m_c,  s)+ \frac{3 \pi}{\alpha^2}\,  \kappa\,
         \sum\limits_{V_i = \psi(1s),\psi(2s)}
      \frac{\Gamma(V_i \rightarrow l^+ l^-)\, m_{V_i}}
{  {m_{V_i}}^2 - q^2  - i m_{V_i} \Gamma_{V_i}}
\right\}
 \nn
&-& \frac12 h(1,s) \left( 4 C_3 + 4 C_4 +3 C_5 + C_6\right)
\nn
&-& \frac12 h(0,s) \left( C_3 + 3 C_4 \right) +
\frac{2}{9} \left( 3 C_3 + C_4 + 3 C_5 + C_6 \right)
\label{eq:C9}
\ena
where $\tilde m_c=\hat m_c/m_1$, $s=q^2/m_1^2$, 
$C_0\equiv 3 C_1 + C_2 + 3 C_3 + C_4+ 3 C_5 + C_6$ and  $\kappa=1/C_0$.
Here
\bea
h(z,s) & = & - \frac{8}{9}\ln\frac{\hat m_b}{\mu}
- \frac{8}{9}\ln z +
\frac{8}{27} + \frac{4}{9} x 
\nn
& - & \frac{2}{9} (2+x) |1-x|^{1/2} \left\{
\begin{array}{ll}
\left( \ln\left| \frac{\sqrt{1-x} + 1}{\sqrt{1-x} - 1}\right| - i\pi
\right), &
\mbox{for } x \equiv \frac{4 z^2}{s} < 1 
\nn
 & \\
2 \arctan \frac{1}{\sqrt{x-1}}, & 
\mbox{for } x \equiv \frac{4 z^2}{s} > 1,
\end{array}
\right. 
\nn
h(0,s) & = & \frac{8}{27} -\frac{8}{9} \ln\frac{\hat m_b}{\mu} -
\frac{4}{9} \ln s + \frac{4}{9} i\pi.
\nonumber
\ena
where $\mu$ is a scale parameter  and $m_1\equiv m_B$.
In what follows, we will not include the long-distance contributions
coming from the $J/\psi$ and $\psi(2S)$ resonances \cite{Ali:1991is}
and charm-loop effects \cite{Khodjamirian:2010vf}.

We specify our choice of the momenta as $p_1=p_2+k_1+k_2$ with 
$p_1^2=m_1^2$, $p_2^2=m_2^2$ and  $k_1^2=k_2^2=m_\ell^2$ where $k_1$ and 
$k_2$ are the $\ell^+$ and $\ell^-$ momenta, and $m_1$, $m_2$, $m_\ell$ are 
the masses of the initial meson $H_1$, the final meson $H_2$, 
and the lepton $\ell$, respectively.
The matrix elements of the exclusive transitions 
$B\to K(K^\ast_0) \bar \ell \ell$  are defined by 
\bea
M(H_1\to H_2 \bar \ell \ell) & = & 
\frac{G_F}{\sqrt{2}}\cdot\frac{ \alpha\lambda_t}{2\,\pi} \cdot 
\left\{C_9^{\rm eff}\,
<H_2\,|\,\bar{s}\,O^\mu\, b\,|\,H_1> \,  \bar \ell\gamma_\mu \ell
\right.
\nn
&+&
C_{10}\, <H_2\,|\,\bar{s}\,O^\mu\, b\,|\,H_1>
\, \bar \ell \gamma_\mu \gamma_5 \ell
\nn
&-& \left.
\frac{2\hat m_b}{q^2}\,C_7^{\rm eff}\, 
<H_2\,|\,  \bar{s}\,i\sigma^{\mu \nu}\,(1+\gamma^5)\, q^\nu \,b\,|\,H_1> \,  
\bar \ell\gamma_\mu \ell 
\right\}\,
\label{eq:mat_el}
\ena
where $H_1 = B$, $H_2 = K (K^\ast_0)$.
 
We define dimensionless form factors by
\bea
< H_2(p_2)\,|\,\bar s\,O^\mu\, b\,| H_1(p_1)> 
&=& 
F_+(q^2)\, P^\mu + F_-(q^2)\, q^\mu\,,
\nn[1.5ex]
< H_2(p_2)\,|\,\bar s \,i\sigma^{\mu\nu}q_\nu(1+\gamma^5)\, b\,|\,H_1(p_1)> 
&=&
- \frac{1}{m_1+m_2} \, \left( P_\mu\, q^2 -  q_\mu\, Pq \right)\, F_T(q^2)\,,
\label{eq:def_ff}
\ena
where $P=p_1+p_2$ and  $q=p_1-p_2$. 
The matrix element in Eq~(\ref{eq:mat_el}) is written as
\[
M\left(H_1\to H_2 + \bar \ell \ell\right)=
\frac{G_F}{\sqrt{2}}\cdot\frac{\alpha\lambda_t}{2\pi}\,
\left\{
T_1^\mu\,(\bar \ell\gamma_\mu \ell)+T_2^\mu\,(\bar \ell\gamma_\mu\gamma_5 \ell)
\right\} 
\]
where the quantities $T_i^\mu$ are expressed through the form
factors and the Wilson coefficients in the case of the spinless
particle $H_2$ as
\bea
T_i^\mu &=& {\cal F}_+^{(i)}\,P^\mu+ {\cal F}_-^{(i)}\,q^\mu \qquad  (i=1,2)\,,
\nn[1.5ex]
{\cal F}_+^{(1)} &=& C_9^{\rm eff}\,F_+ + C_7^{\rm eff}\,F_T\, 
\frac{2\hat m_b}{m_1+m_2}\,,
\nn[1.5ex]
{\cal F}_-^{(1)} &=& C_9^{\rm eff}\,F_- - C_7^{\rm eff}\,F_T\,
\frac{2\hat m_b}{m_1+m_2}\,\frac{Pq}{q^2}\,,
\nn[1.5ex]
{\cal F}_\pm^{(2)} &=& C_{10}\,F_\pm\,.
\label{eq:amp_pp}
\ena
Respectively, the helicity form factors $H^i_m$ are defined in terms of the
invariant form factors as \cite{Faessler:2002ut}
\be
H^i_t = \frac{1}{\sqrt{q^2}}(Pq\, {\cal F}^i_+ + q^2\, {\cal F}^i_-)\,,
\qquad
H^i_\pm = 0\,,
\qquad
H^i_0  = \frac{2\,m_1\,|{\bf p_2}|}{\sqrt{q^2}} \,{\cal F}^i_+ \,.
\label{eq:hel_pp}
\en 

The differential $(q^2,\cos\theta)$ two-fold decay distribution may be written 
in terms of the bilinear combinations of the helicity amplitudes
(see Ref.~\cite{Faessler:2002ut}). However, it is common in the modern
literature to use the transversality amplitudes $A^{L,R}_{\perp,\parallel,0}$ 
and $A_t$ defined in Ref.~\cite{Kruger:2005ep}.
They are related to our helicity amplitudes by
\bea
A^{L,R}_\perp &=& N \,\frac{1}{\sqrt{2}}
\left[ (H^{(1)}_+ -  H^{(1)}_-) \mp  (H^{(2)}_+ -  H^{(2)}_-) \right]\,,
\nn[1.2ex]
A^{L,R}_\parallel &=& N \,\frac{1}{\sqrt{2}}
\left[ (H^{(1)}_+ +  H^{(1)}_-) \mp  (H^{(2)}_+ +  H^{(2)}_-) \right]\,,
\nn[1.2ex]
A^{L,R}_0 &=& N\,\left( H^{(1)}_0 \mp   H^{(2)}_0 \right)\,,
\nn[1.2ex]
A_t &=& -2\,N\,H^{(2)}_t
\label{eq:eq:KGvsOur}
\ena
where the overall factor is given by
\[
N = \Big[ \frac14 \frac{G_F^2}{(2\pi)^3} 
                   \left( \frac{\alpha \lambda_t}{2\pi} \right)^2
                  \frac{ {\bf|p_2|} q^2 v }{12 m_1^2 } 
     \Big]^{\frac12} 
\]
where ${\bf|p_2|}=\lambda^{1/2}(m_1^2,m_2^2,q^2)/2m_1$ is the momentum
of the outgoing meson $H_2$ and $v= \sqrt{1-4m^2_\ell/q^2}$ is the lepton
velocity, both of which are given in the rest frame of the parent meson $H_1$.

The differential decay distribution then reads
\bea
\frac{d\Gamma(H_1\to H_2\bar \ell \ell)}{dq^2d(\cos\theta)} 
&=&\,
\frac{G_F^2}{(2\pi)^3} \left( \frac{\alpha \lambda_t}{2\pi} \right)^2
\frac{ {\bf|p_2|} q^2 v }{12 m_1^2 }
\nn
&\times&\frac{3}{16}
\Big\{
                       |H_0^{(1)}|^2 + |H_0^{(2)}|^2
    +2\,\delta_{\ell\ell} \left[|H_0^{(1)}|^2 - |H_0^{(2)}|^2\right] 
    +4\,\delta_{\ell\ell} |H_t^{(2)}|^2
\nn
&-& \cos2\theta\, (1-2\, \delta_{\ell\ell})
    \left[|H_0^{(1)}|^2 + |H_0^{(2)}|^2\right]
\Big\}
\nn[2ex]
&=& 
\frac38\Big\{
  |A^L_0|^2 + |A^R_0|^2 
+ 4\,\delta_{\ell\ell}\,{\rm Re}\left(A^L_0A^{R\,\dagger}_0 \right)
+ 2\,\delta_{\ell\ell}|A_t|^2
\nn
&-& \cos2\theta\, (1-2\, \delta_{\ell\ell}) \left[ |A^L_0|^2 + |A^R_0|^2 \right]
\Big\}\,.
\label{eq:distr2}
\ena
Integrating over $\cos\theta$ one obtains 
\bea
\frac{d\Gamma(H_1 \to H_2 \bar \ell \ell)}{dq^2} 
&=&\,
\frac{G_F^2}{(2\pi)^3} \left( \frac{\alpha \lambda_t}{2\pi} \right)^2
\frac{ {\bf|p_2|} q^2 v }{12 m_1^2 }
\nn
&\times&
\frac{1}{2}
\Big\{
 |H_0^{(1)}|^2 + |H_0^{(2)}|^2
 +\delta_{\ell\ell} \left[|H_0^{(1)}|^2 - 2 |H_0^{(2)}|^2 +3|H_t^{(2)}|^2 \right] 
\Big\}
\nn[2ex]
&=&
 (1-\tfrac12 \delta_{\ell\ell}) \left[|A^L_0|^2 + |A^R_0|^2 \right]
 + \tfrac32\,\delta_{\ell\ell} \left[ 2\,{\rm Re}\left(A^L_0A^{R\,\dagger}_0 \right)
                                   +|A_t|^2 \right]
\label{eq:distr1}
\ena
where we have introduced a flip parameter $\delta_{\ell\ell}= 2m^2_\ell/q^2$.

We also calculate the differential rates for the semileptonic
$H_1\to H_2 \ell\bar\nu_\ell$ mode and rare $H_1\to H_2\nu\bar\nu$ decay.
One has

\bea
\frac{d\Gamma(H_1 \to H_2 \ell\bar\nu_\ell)}{dq^2} 
&=&\,
\frac{G_F^2}{(2\pi)^3} |V_{bu}|^2
\frac{ {\bf|p_2|} q^2}{12 m_1^2 }\left(1-\frac{m_\ell^2}{q^2}\right)^2
\Big\{
 \left(1+\frac{m_\ell^2}{2q^2}\right) |H_0|^2 
+ \frac{3m_\ell^2}{2q^2}|H_t|^2
\Big\}\,,
\label{eq:semilep}\\
H_0 &=& \frac{2m_1{\bf|p_2|}}{\sqrt{q^2}}\,F_+\,,\qquad
H_t=\frac{1}{\sqrt{q^2}}\left( Pq F_+ + q^2 F_-\right)\,,
\nn[2ex]
\frac{d\Gamma(H_1 \to H_2 \nu\bar\nu)}{dq^2} 
&=&\,
\frac{G_F^2}{(2\pi)^3} \left( \frac{\alpha \lambda_t}{2\pi} \right)^2
\frac{ {\bf|p_2|}^3}{\sin^4\theta_W} |D_\nu(x_t)|^2 F^2_+(q^2)\,.
\label{eq:neutrino}
\ena
The form factors $F_\pm$ are defined by Eq.~(\ref{eq:def_ff}),
whereas the function $D_\nu(x_t)$ is given by
\[
D_\nu(x_t) = \frac{x_t}{8}\left(\frac{2+x_t}{x_t-1}
           + \frac{3x_t-6}{(x_t-1)^2}\ln x_t\right)\,, \qquad
x_t=\frac{\hat m_t^2}{m_W^2}\,.
\]

\section{The $B-S$ transition form factors in the covariant
quark model}

We calculate the $B-S$ transition form factors in the covariant
quark model. We briefly recall the basic features of this approach,
which was formulated in its modern form in Ref.~\cite{Branz:2009cd} 
by taking into account the infrared confinement of quarks. 

The model is based on an effective interaction Lagrangian describing the
coupling of hadrons to their constituent quarks.
For instance, the coupling of a meson $M(q_1 \bar q_2)$ to its constituent 
quarks $q_1$ and $\bar q_2$ is described by the nonlocal Lagrangian  
\be
\label{eq:lag}
{\cal L}_{\rm int}^{\rm str}(x) = g_M M (x) \, 
\int\!\! dx_1 \!\! \int\!\! dx_2 
F_M(x,x_1,x_2) \,
\bar q_1(x_1) \, \Gamma_M \,  q_2(x_2)  \, + \, {\rm H.c.}  
\en 
Here, $\Gamma_M$ is  the Dirac matrix, which is chosen appropriately 
to describe the spin quantum numbers of the meson field $M(x)$.
The vertex function $F_M(x,x_1,x_2)$ characterizes 
the finite size of the meson.  To satisfy translational invariance the 
vertex function has to obey the identity 
$F_M(x+a,x_1+a,x_2+a) \, = \, F_M(x,x_1,x_2) $
for any given four-vector $a$. 
We use a specific form for the vertex function which 
satisfies the above translation invariance relation. One has 
\be
\label{eq:vertex}
F_M(x,x_1,x_2) \, = \, \delta^{(4)}(x - \sum\limits_{i=1}^2 w_i x_i) \;  
\Phi_M\biggl( (x_1 - x_2)^2 \biggr) 
\en
where $\Phi_M$ is the correlation function of the two constituent quarks 
with masses $m_1$ and $m_2$. 
The variable $w_i$ is defined by $w_i=m_i/(m_1+m_2)$, so that 
$w_1+w_2=1$. 
We choose a simple Gaussian form for the vertex function $\Phi_M(-k^2)$. 
The minus sign in the argument of $\Phi_{M}(-k^2)$ is chosen to emphasize
that we are working in Minkowski space.
One has 
\be
\label{eq:fourier}
\widetilde\Phi_{M}(-k^2) = \exp( k^2/\Lambda_M^2) 
\en 
where the parameter $\Lambda_M$ characterizes the size of the meson. 
Since $k^2$ turns into $-k^2_E$ in  Euclidean space 
the form~(\ref{eq:fourier}) has the appropriate falloff
behavior in the Euclidean region. We stress that any choice for 
$\Phi_M$ is appropriate as long as it falls off sufficiently fast 
in the ultraviolet region of Euclidean space in order to render 
the Feynman diagrams ultraviolet finite. 

In the evaluation of the quark-loop diagrams we use
the free local fermion propagator for the constituent quark,
\be
\label{eq:prop}
S_q(k) = \frac{1}{ m_q-\not\! k -i\epsilon }
\en 
with an effective constituent quark mass $m_q$. 

\begin{figure}[ht]
\begin{center}
\includegraphics[width=0.6\textwidth]{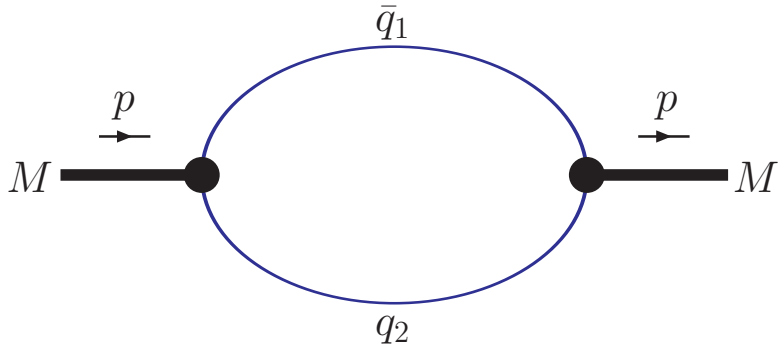}
\caption{Diagram describing the meson mass operator.}
\label{fig:mass}
\end{center}
\end{figure}

The coupling constant $g_M$ in Eq.~(\ref{eq:lag}) is determined by the 
so-called {\it compositeness condition}  suggested by 
Weinberg~\cite{Weinberg:1962hj} and Salam~\cite{Salam:1962ap} 
(for a review, see Ref.~\cite{Hayashi:1967hk}) and extensively used in 
our studies (for details, see Ref.~\cite{Efimov:1993ei}). 
The compositeness condition requires that the renormalization constant $Z_M$ 
of the elementary meson field $M(x)$ is set to zero, i.e.,
\be
\label{eq:Z=0}
Z_M=1-g_M^2\Pi^\prime_M(m_M^2)=0
\en
where $\Pi^\prime_M(p^2)$ is the derivative of the mass operator
corresponding to the self--energy diagram  in Fig.~\ref{fig:mass}. 

To clarify the physical meaning of the compositeness condition, we recall 
that the renormalization constant $Z_M^{1/2}$ can also be interpreted as  
the matrix element between the physical state and the corresponding 
bare state.  For $Z_M=0$ it then follows that the physical state does not 
contain  the bare one and it is therefore described as a bound state. 
The interaction Lagrangian (\ref{eq:lag}) and 
the corresponding free Lagrangian describe  
both the constituents (quarks) and the physical particles (hadrons), 
which are bound states of the constituents.
As a result of the interaction, the physical particle is dressed, 
i.e., its mass and wave function have to be renormalized. 
The condition $Z_M=0$ also effectively excludes
the constituent degrees of freedom from the space of physical states
and thereby guarantees that there will be no double counting.
The constituents exist in virtual states only. 

The covariant quark model was applied  to evaluate the form factors
of the $B(B_s)\to P(V)-$transitions in the full kinematical
region of momentum transfer squared \cite{Ivanov:2011aa,Dubnicka:2013vm}
This approach was extended to describe the baryons as three-quark states
\cite{baryon} and the exotic meson X(3872) as a tetraquark \cite{tetraquark}.

A similar approach based on the compositeness condition $Z=0$ was recently
developed in Ref.~\cite{Cheung:2014cka}.

In this paper we evaluate the $B-S$ transition form factors
assuming that the scalar mesons below 1 GeV are ordinary two-quark states.
Some remarks should be made before performing the calculations.
The internal structure of the light scalar mesons is not yet well established
(for review, see Refs.~\cite{Amsler:2013wea,Amsler:2004ps}).
Since they have large decay widths it is difficult to distinguish them
from background. There are interpretations of these objects as
four-quark states and/or gluballs. Here, we describe the scalar mesons
as two-quark states and evaluate the $B-S$ form factors within
our approach, but when we use the calculated form factors in
the matrix element of the cascade decay $B\to K^\ast_0(\to K\pi)\ell^+\ell^-$
we take into account the line shape of the $K^\ast_0$, which reflects
the broad width of this resonance. One can also describe the scalar
mesons as four-quark states in our approach, similar to the exotic meson 
X(3872) \cite{tetraquark}; however, this is beyond the scope of this work.

The SU(3) nonet of scalar mesons below 1 GeV
can be written in the matrix form

\be
\hat S = \frac{1}{\sqrt{2}}\sum_{i=0}^8 S^i\lambda^i\, \qquad 
\lambda^0=\sqrt{\frac{2}{3}}\,I.
\label{eq:nonet}
\en
The physical scalar fields are related to the Cartesian basis 
in the following manner:

\be
\begin{array}{ll}
S^\pm =\frac{1}{\sqrt{2}}(S^1 \mp i S^2)\,, \qquad & S^0=S^3 \,,
\\[1.5ex]
S^+_s = \frac{1}{\sqrt{2}}(S^4 - i S^5)\,, \qquad  &
S^0_s =\frac{1}{\sqrt{2}}(S^6 - i S^7)\,,
\\[1.5ex]
     S^-_s  = \frac{1}{\sqrt{2}}(S^4 + i S^5)\,, \qquad &
\bar S^0_s =\frac{1}{\sqrt{2}}(S^6 + i S^7)\,,
\\[1.5ex]
S' =   S^0 \cos\theta_S  + S^8 \sin\theta_S\,, \qquad &
S  = - S^0 \sin\theta_S  + S^8 \cos\theta_S\,,
\\
\end{array}
\label{eq:phys_scalar}
\en
where $\theta_S$ is the octet-singlet mixing angle.
The $\bar q \hat S q$ vertex is then written as
\bea
{\cal L}_{S\bar q q} &=& \bar q \hat S q
\nn
&=& S^+\,\bar ud + S^-\,\bar du + S^0\, \tfrac{1}{\sqrt{2}}(\bar uu- \bar dd)
+ S_s^+\,\bar us + S_s^0\,\bar ds +  S_s^-\,\bar su + \bar S_s^0\,\bar sd
\nn
&+& S'\,\Big(  \cos\delta_S \tfrac{1}{\sqrt{2}}(\bar uu + \bar dd)
        - \sin\delta_S\, \bar ss \Big)
- S\,\Big(  \sin\delta_S \tfrac{1}{\sqrt{2}}(\bar uu + \bar dd)
        + \cos\delta_S\, \bar ss \Big)\,,
\label{eq:Sqq}
\ena 
where $\delta_S=\theta-\theta_I$, with the ideal mixing angle
$\theta_I=\arctan\left(1/\sqrt{2}\right)$.
We will use the notation from Ref.~\cite{Agashe:2014kda}
for the scalar mesons below 1~GeV:
\begin{itemize}
\item $S_s\equiv K^\ast_0(800)$,\,\,  $I\,(J^{P})=\frac12(0^{+})$,\,\, 
$m_{ K^\ast_0(800)}= 682\pm 29 $~MeV;
\item $S'\equiv f_0(500)$,\,\,  $I^G\,(J^{PC})=0^+(0^{++})$, \,\,
$m_{f_0(500)}= 400-550 $~MeV;
\item $S\equiv f_0(980)$,\,\,  $I^G(J^{PC})=0^+(0^{++})$, \,\,
$m_{f_0(980)}= 990\pm 20 $~MeV;
\item $S^{\pm,0}\equiv a^{\pm,0}_0(980)$,\,\,  $I^G(J^{PC})=1^-(0^{++})$, \,\,
$m_{a_0(980)}= 980\pm 20 $~MeV.
\end{itemize}
Moreover, we assume that $\delta_S=0$, i.e., $m_{f_0(980)}$ to ensure a pure
$\bar ss$ state. 

The coupling constant $g_S$ in Eq.~(\ref{eq:lag}) is determined by 
Eq.~(\ref{eq:Z=0}),
where $\widetilde\Pi^\prime_S$ is the derivative of the scalar meson mass 
operator, 

\bea\label{eq:Mass-operator}
\widetilde\Pi'_S(p^2) &=&
-\,\frac{1}{2p^2}\,
p^\alpha\frac{d}{dp^\alpha}\,
\int\!\! \frac{d^4k}{4\pi^2i}\, \widetilde\Phi^2_S(-k^2)\,
{\rm tr} \biggl[S_1(k+w_1 p)\,S_2(k-w_2 p) \biggr] 
\nn
&=&
-\,\frac{1}{2p^2}\,\int\!\! \frac{d^4k}{4\pi^2i}\,\widetilde\Phi^2_S(-k^2)\,
\Big\{
w_1\,{\rm tr} 
\biggl[S_1(k+w_1 p)\!\not\!p\, S_1(k+w_1 p)\,S_2(k-w_2 p) \biggr] 
\nn
&&
\phantom{
\frac{1}{2p^2}\,\int\!\! \frac{d^4k}{4\pi^2i}\,\widetilde\Phi^2_S(-k^2)\,
\!\!\!
}
-w_2\,{\rm tr} 
\biggl[S_1(k+w_1 p)\,S_2(k-w_2 p)\!\not\!p\, S_2(k-w_2 p) \biggr]
\Big\}\,. 
\ena

By using the calculation technique outlined in Ref.~\cite{Branz:2009cd},
one can easily perform the loop integration. We give the analytic result
for equal quark masses ($m_{q_1}=m_{q_2}\equiv m_q$):
\bea
\tilde\Pi'_S(p^2)&=& \int\limits_0^{1/\lambda^2} \!\! 
\frac{dt\,t}{a_S^2} \int\limits_0^1\!\!d\alpha\,
e^{-t\,z_0 + z_1}\,
\nn
&\times&
\frac{t}{32}
\Big\{
 p^2-4 m^2_q + \frac{1}{a_S} \left[ 20 +t\,(1-2\alpha)^2 (12 m^2_q-p^2)\right]
            - \frac{t}{a^2_S} (1-2\alpha)^2 (12 + p^2 t)
            + \frac{t^3}{a^3_S} (1-2\alpha)^4 p^2
\Big\}
\label{eq:mass_fin}\\[2ex]
z_0 &=&  \alpha m^2_q - \alpha(1-\alpha) p^2 ,
\qquad z_1  = \frac{st}{2a_S} (1-2\alpha)^2 p^2 ,
\nn[2ex] 
a_S &=& 2s+t\, , \qquad s = \frac{1}{\Lambda_S^2}\,. 
\nonumber
\ena
Note that in the case of $\lambda\to 0$ the branching point appears 
at $p^2=4m_q^2$. At this point the integral over $t$ becomes 
divergent as $t\to\infty$ because  $z_0=0$
at $\alpha=1/2$.  By introducing an infrared cutoff 
on the upper limit of the scale of integration, one can avoid the appearance 
of the threshold singularity.

\begin{figure}[ht]
\begin{center}
\includegraphics[width=0.60\textwidth]{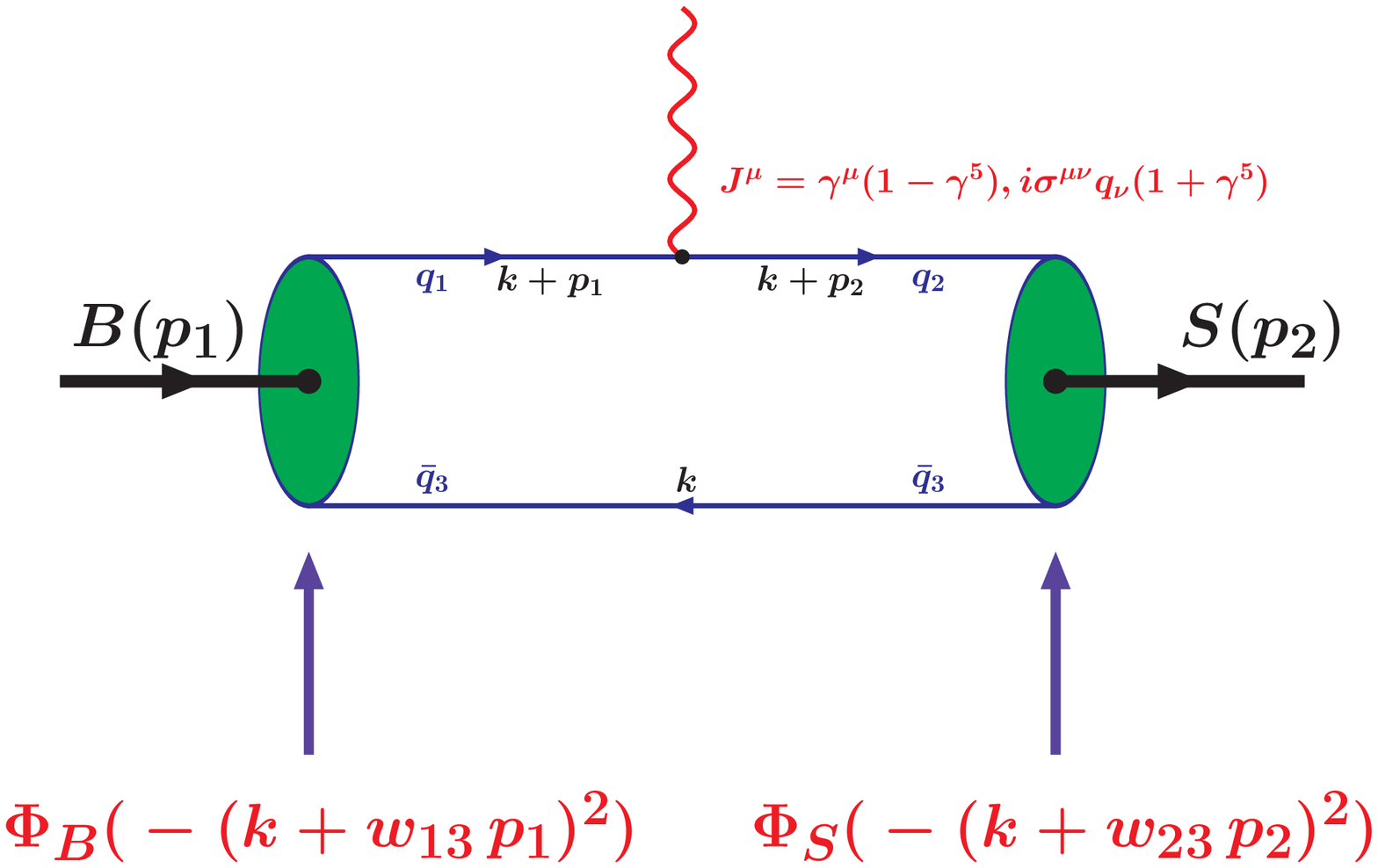} 
\caption{Diagrammatic representation of the matrix elements
in Eqs.~(\ref{eq:BS}) and (\ref{eq:BST}).
}
\label{fig:diag}
\end{center}
\end{figure}

Herein our primary subjects are the $B-S$ transition matrix elements, 
which can be expressed  via the dimensionless form factors
defined in Refs.~\cite{Ivanov:2011aa,Dubnicka:2013vm}.
The diagram corresponding to these matrix elements is shown
in Fig.~\ref{fig:diag}.

One has
\bea
&&
\langle 
S_{[\bar q_3 q_2]}(p_2)\,|\,\bar q_2\, O^{\,\mu}\, q_1\,| B_{[\bar q_1 q_3]}(p_1)
\rangle
\nn
&=&
N_c\, g_B\,g_S\!\!  \int\!\! \frac{d^4k}{ (2\pi)^4 i}\, 
\widetilde\Phi_B\Big(-(k+w_{13}p_1)^2\Big)\,
\widetilde\Phi_S\Big(-(k+w_{23}p_2)^2\Big)
\nn
&\times&
{\rm tr} \biggl[
S_2(k+p_2) O^{\,\mu}\, S_1(k+p_1)\, \gamma^5\, S_3(k)\, 
\biggr]
\nn
 & = & F^{BS}_+(q^2)\, P^{\,\mu} + F^{BS}_-(q^2)\, q^{\,\mu}\,,
\label{eq:BS}\\[2ex]
&&
\langle 
S_{[\bar q_3 q_2]}(p_2)\,|
\,\bar q_2\, (i\sigma^{\,\mu\nu}q_\nu(1+\gamma^5)) \, q_1\,
                     | B_{[\bar q_1 q_3]}(p_1)
\rangle
\nn
&=&
N_c\, g_B\,g_S\!\!  \int\!\! \frac{d^4k}{ (2\pi)^4 i}\, 
\widetilde\Phi_B\Big(-(k+w_{13}p_1)^2\Big)\,
\widetilde\Phi_S\Big(-(k+w_{23}p_2)^2\Big)
\nn
&\times&
{\rm tr} \biggl[
S_2(k+p_2) i\sigma^{\,\mu\nu}q_\nu(1+\gamma^5)\, S_1(k+p_1)\, \gamma^5\, S_3(k)\, 
\biggr]
\nn
 & = & 
-\,\frac{1}{m_1+m_2}\,\left(q^2\, P^{\,\mu}-q\cdot P\, q^{\,\mu}\right)\,
F^{BS}_T(q^2).
\label{eq:BST}
\ena
Here, $p_i^2=m_i^2$, $q_1=b$, $q_2=u,s,d$, and $q_3=s,d$. 
Since there are three sorts of quarks involved in these
processes, we introduce the notation with two subscripts,
$w_{ij}=m_{q_j}/(m_{q_i}+m_{q_j})$ $(i,j=1,2,3)$ so that $w_{ij}+w_{ji}=1$. 

The first fit of the model parameters was done in
the original paper \cite{Branz:2009cd}, where the infrared quark
confinement was implemented for the first time. 
The leptonic decay constants (which are known either from experiments
or from lattice simulations)  have been chosen as the input quantities
to adjust the model parameters. A given meson $H$ in the interaction
Lagrangian is characterized by the coupling constant $g_H$,
the size parameter $\Lambda_H$ and two of the four constituent quark masses,
$m_q$ ($m_u=m_d$, $m_s$, $m_c$, $m_b$). Moreover, there is the infrared
confinement parameter $\lambda$, which is universal for all hadrons.
Note that the physical values for the hadron masses have been used
in the fit. Therefore, one  has $2n_H+5$ adjustable parameters for $n_H$ 
numbers of mesons. The compositeness condition provides 
$n_H$ constraints and allows one to express all coupling constants $g_H$
via other model parameters. The remaining  $n_H+5$ parameters are determined by
a fit to experimental data.  The values of leptonic decay 
constants and some electromagnetic decay widths have been chosen as the input
data. Several updated fits were done in 
Refs.~\cite{Ivanov:2011aa,Dubnicka:2013vm}.
In this paper we will use the latest fit done in Ref.~\cite{Ganbold:2014pua}.
The  fitted values of the constituent quark masses $m_q$, the infrared cut-off
$\lambda$, and the size parameters $\Lambda_H$ are given by
Eq.~(\ref{eq:fitmas}) and Table~\ref{tab:Lambda-H}. 

\be
\def\arraystretch{1.5}
\begin{array}{cccccc}
     m_{u/d}        &      m_s        &      m_c       &     m_b & \lambda  &   
\\\hline
 \ \ 0.241\ \   &  \ \ 0.428\ \   &  \ \ 1.67\ \   &  \ \ 5.05\ \   & 
\ \ 0.181\ \   & \ {\rm GeV} 
\end{array}
\label{eq:fitmas}
\en

\begin{table}[ht]
\caption{The fitted values of the size parameters  $\Lambda_H$ in GeV.}
\label{tab:Lambda-H}
\begin{center}
\def\arraystretch{1.5}
\begin{tabular}{cccccccccc}
\hline
 $\pi$ & $K$  & $D$  & $D_s$ & $B$  & $B_s$ & $B_c$ & $\eta_c$ & $\eta_b$ &\\ 
%\hline
  0.87 & 1.02 & 1.71 & 1.81  & 1.96 & 2.05  & 2.50 & 2.06 & 2.95 &\\
\hline\hline
 $\rho$ & $\omega$ & $\phi$ & $J/\psi$ & $K^\ast$ & $D^\ast$ & $D_s^\ast$ & 
  $B^\ast$ & $B_s^\ast$ & $\Upsilon$ \\ 
%\hline
 \ \ 0.61\ \  & \ \  0.50 \ \  & \ \  0.91\ \  & \ \  1.93\ \  & \ \  0.75\ \  
 & \ \  1.51\ \  & \ \  1.71 \ \  & \ \  1.76\ \  & \ \  1.71 \ \ & 
\ \ 2.96 \ \ \\ 
\hline
\end{tabular}
\end{center}
\end{table}

Our form factors are represented as three-fold integrals which
are calculated by using NAG routines. 
The results of our numerical calculations are well approximated
by the parametrization
\be
F(q^2)=\frac{F(0)}{1-a s+b s^2}\,, \qquad s=\frac{q^2}{m_1^2}\,.
\label{eq:ff_approx}
\en
We consider the following weak transitions: $b-u$ (charged current), and
$b-d$ and $b-s$ (flavor-changing neutral currents).
The values of $F(0)$, $a$, and $b$ are listed  in Table~\ref{tab:ff}. 

\begin{table}[ht]
\begin{center}
\def\arraystretch{1.5}
\begin{tabular}{|c|l|c|c|c|c|c|c|}
\hline
$q_1-q_2$ & \qquad $B-S$  & \multicolumn{3}{c|}{ $\Lambda_S=0.8$}  &\multicolumn{3}{c|} {$\Lambda_S=1.5$} \\
\cline{3-8}
      &                          & $F_{+}(0)$ & $a_{+}$ & $b_{+}$   & $F_{+}(0)$ & $a_{+}$ & $b_{+}$     \\
\hline
$b-u$ & $B^0_d-a^+_0(980)$        & 0.144     & 1.624   & 0.585    & 0.192     & 1.433  & 0.381 \\
\hline
$b-u$ & $B^0_s-K^{\ast\,+}_0(800)$  & 0.138     & 1.667   & 0.674    & 0.274     & 1.258  & 0.292  \\
\hline
$b-s$ & $B^0_s-f_0(980) $         & 0.141     & 1.663   & 0.651    & 0.254     & 1.269  & 0.262 \\
\hline
$b-s$ & $B^0_d-K^{\ast\,0}_0(800)$  & 0.191     & 1.348   & 0.407    & 0.306     & 0.988  & 0.108  \\
\hline
$b-d$ & $B^0_d-f_0(500) $         & 0.120     & 1.448   & 0.485    & 0.210     & 1.067  & 0.155  \\
\hline
\end{tabular}
\end{center}
\begin{center}
\def\arraystretch{1.5}
\begin{tabular}{|c|l|c|c|c|c|c|c|}
\hline
 $q_1-q_2$ & \qquad $B-S$ & \multicolumn{3}{c|}{ $\Lambda_S=0.8$}  &\multicolumn{3}{c|} {$\Lambda_S=1.5$} \\
\cline{3-8}
      &                         & $-F_{-}(0)$  & $a_{-}$  & $b_{-}$ & $-F_{-}(0)$ & $a_{-}$ & $b_{-}$ \\
\hline
$b-u$ & $B^0_d-a^+_0(980) $      & 0.049       & 2.144  & 1.196    & 0.089   & 1.723     & 0.688 \\
\hline
$b-u$ & $B^0_s-K^{\ast\,+}_0(800)$ & 0.138       & 1.727  & 0.734    & 0.268   & 1.291     & 0.310  \\
\hline
$b-s$ & $B^0_s-f_0(980)$         & 0.140       & 1.761  & 0.755    & 0.253   & 1.320     & 0.295  \\
\hline
$b-s$ & $B^0_d-K^{\ast\,0}_0(800)$ & 0.199       & 1.406  & 0.457    & 0.296   & 1.032     & 0.129  \\
\hline
$b-d$ & $B^0_d-f_0(500)$         & 0.116       & 1.504  & 0.536    & 0.191   & 1.110     & 0.180 \\
\hline
\end{tabular}
\end{center}
\begin{center}
\def\arraystretch{1.5}
\begin{tabular}{|c|l|c|c|c|c|c|c|}
\hline
 $q_1-q_2$ & \qquad $B-S$        & \multicolumn{3}{c|}{ $\Lambda_S=0.8$}  &\multicolumn{3}{c|} {$\Lambda_S=1.5$} \\
\cline{3-8}
          &                     & $F_{T}(0)$ & $a_{T}$ & $b_{T}$ & $F_{T}(0)$ & $a_{T}$  & $b_{T}$ \\
\hline
%$b-u$ & $B^0_d-a^+_0(980)$       & 0.134     & 1.714  & 0.685   & 0.182     & 1.499    & 0.445 \\
%\hline
%$b-u$ & $B^0_s-K^{\ast\,+}_0(800)$ & 0.154     & 1.666 & 0.671    & 0.292     & 1.251    & 0.277 \\
%\hline
$b-s$ & $B^0_s-f_0(980) $        & 0.165     & 1.680 & 0.667    & 0.285     & 1.276    &  0.257 \\ 
\hline
$b-s$ & $B^0_d-K^{\ast\,0}_0(800)$ & 0.206     & 1.367 & 0.423    & 0.306     & 1.005    & 0.113 \\
\hline
$b-d$ & $B^0_d-f_0(500) $        & 0.124     & 1.460 & 0.496    & 0.203     & 1.080    & 0.159 \\
\hline
\end{tabular}
\caption{ The parameters of the fitted  transition form factors $F(q^2)$ 
at $\Lambda_S = 0.8 $ GeV 
  and $\Lambda_S = 1.5 $ GeV.  }
\label{tab:ff}
\end{center}
\end{table}

\section{Numerical results and discussion}

We use the following set of SM parameters:
$G_F = 1.16637\times 10^{-5}$~GeV$^{-2}$, 
$\hat m_c=1.27$~GeV, $\hat m_b=4.19$~GeV,
$\hat m_t=173.8$~GeV, $m_W=80.41$~GeV, $\sin^2\theta_W=0.2233$,
$\lambda_t=|V_{ts}V^\dagger_{tb}|=0.041$, $|V_{ub}|=0.00413$, 
a scale parameter $\mu=\hat m_b$, and the Wilson coefficients
$C_1 = -0.248$, $C_2 = 1.107$, $C_3 = 0.011$, $C_4 = -0.026$,
$C_5 = 0.007$, $C_6 = -0.031$, $C_7^{\rm eff}==-0.313$, $C_9 = 4.344$, and
$C_{10} = -4.669$. We take the average values of the hadron and lepton masses
 and the $B_d(B_s)$-meson lifetimes from Ref.~\cite{Agashe:2014kda}.

All model parameters are fixed by fitting the experimental data
in our previous papers (see 
Refs.~\cite{Ivanov:2011aa,Dubnicka:2013vm,Ganbold:2014pua,baryon}).
Their numerical values are shown in Eq.~(\ref{eq:fitmas}) and 
Table~\ref{tab:Lambda-H}. The only new parameter is $\Lambda_S$
which characterizes the size of the scalar mesons.  We allow this parameter
to vary in a relatively large interval,  $\Lambda_S\in [0.8,1.5]$~GeV.

\begin{figure}[ht]
\begin{center}
\hspace*{-0.5cm}
\begin{tabular}{lr}
\includegraphics[width=0.40\textwidth]{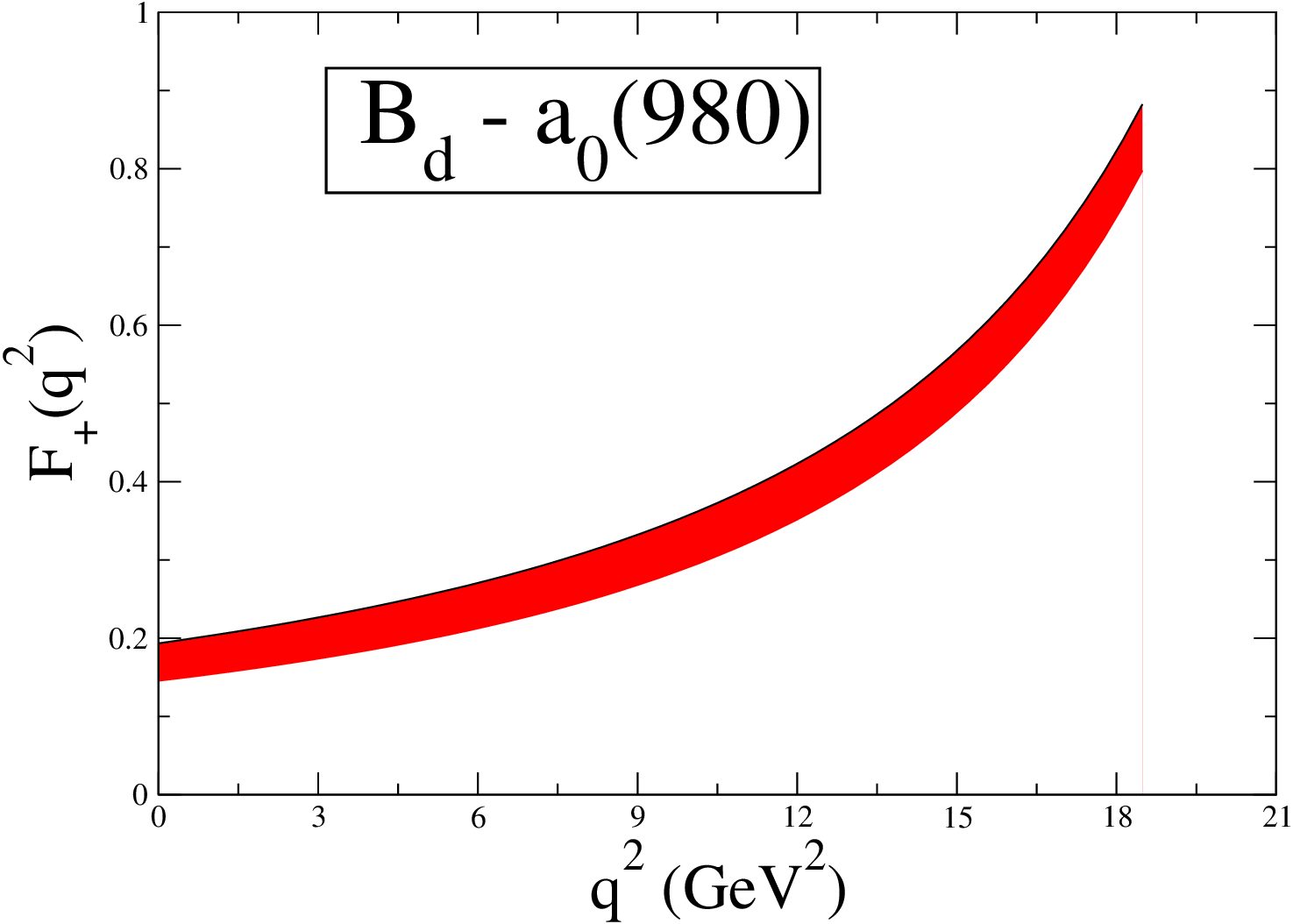}  \qquad &\qquad
\includegraphics[width=0.40\textwidth]{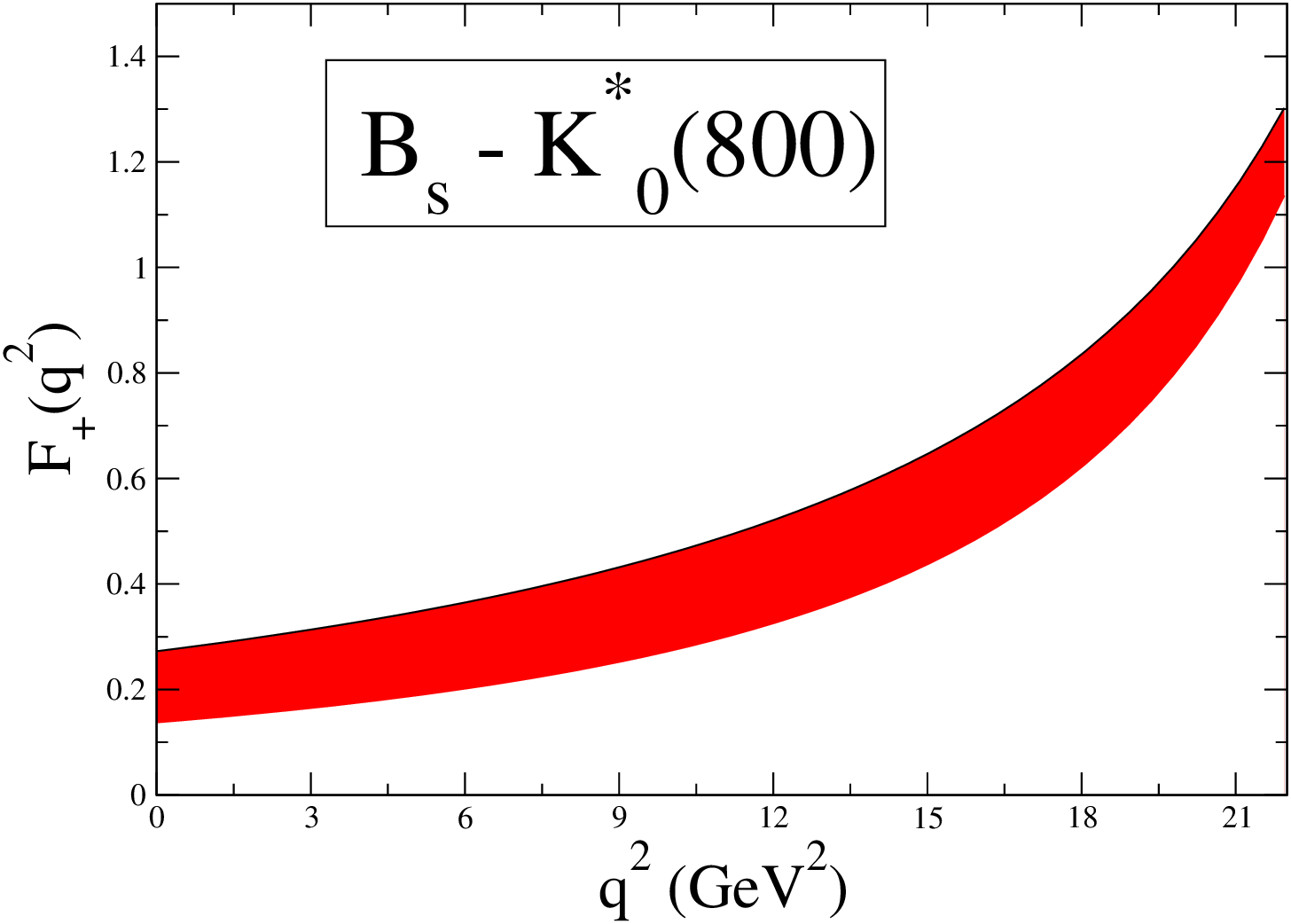} 
\\[6ex]
\includegraphics[width=0.40\textwidth]{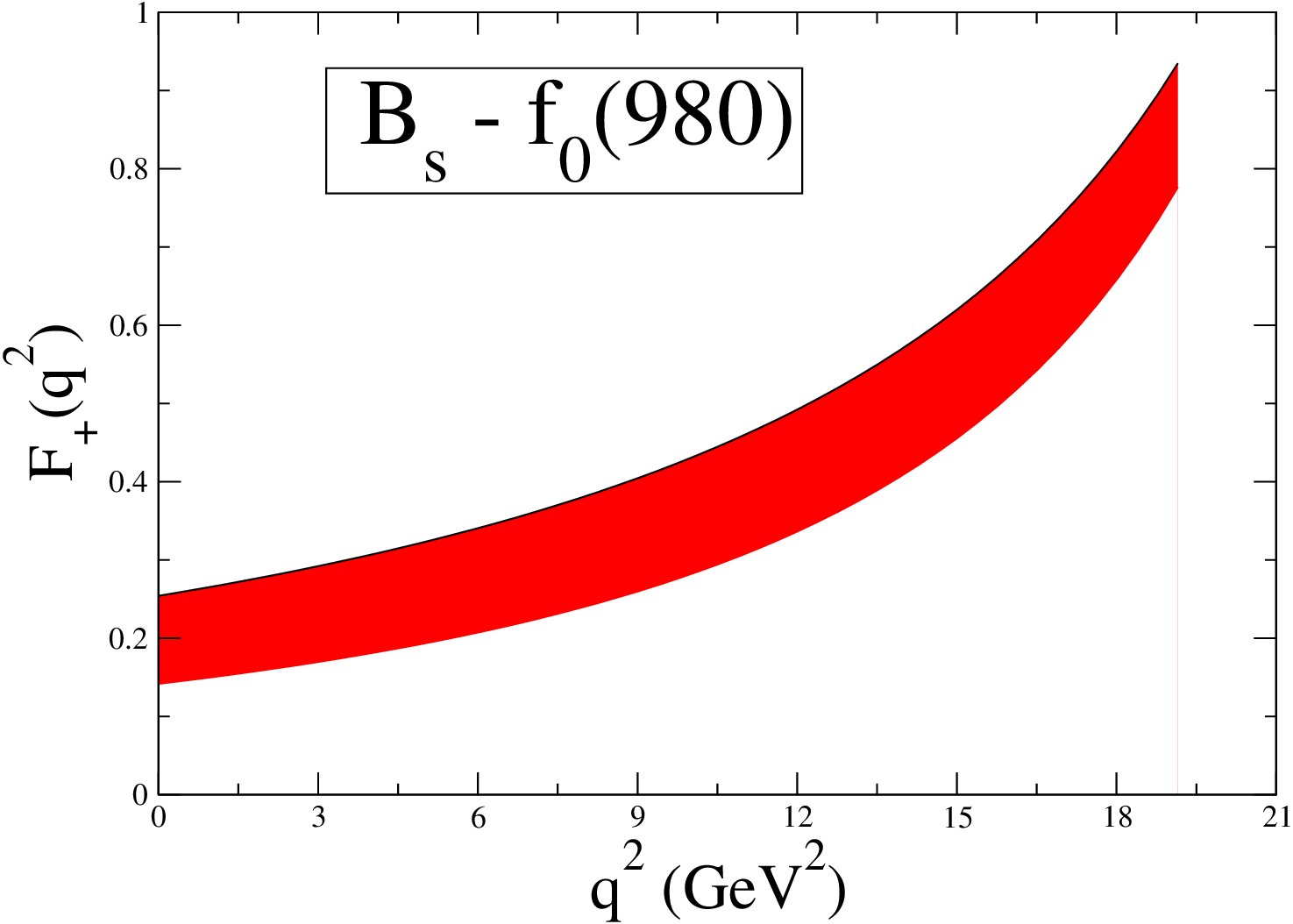}  \qquad & \qquad
\includegraphics[width=0.40\textwidth]{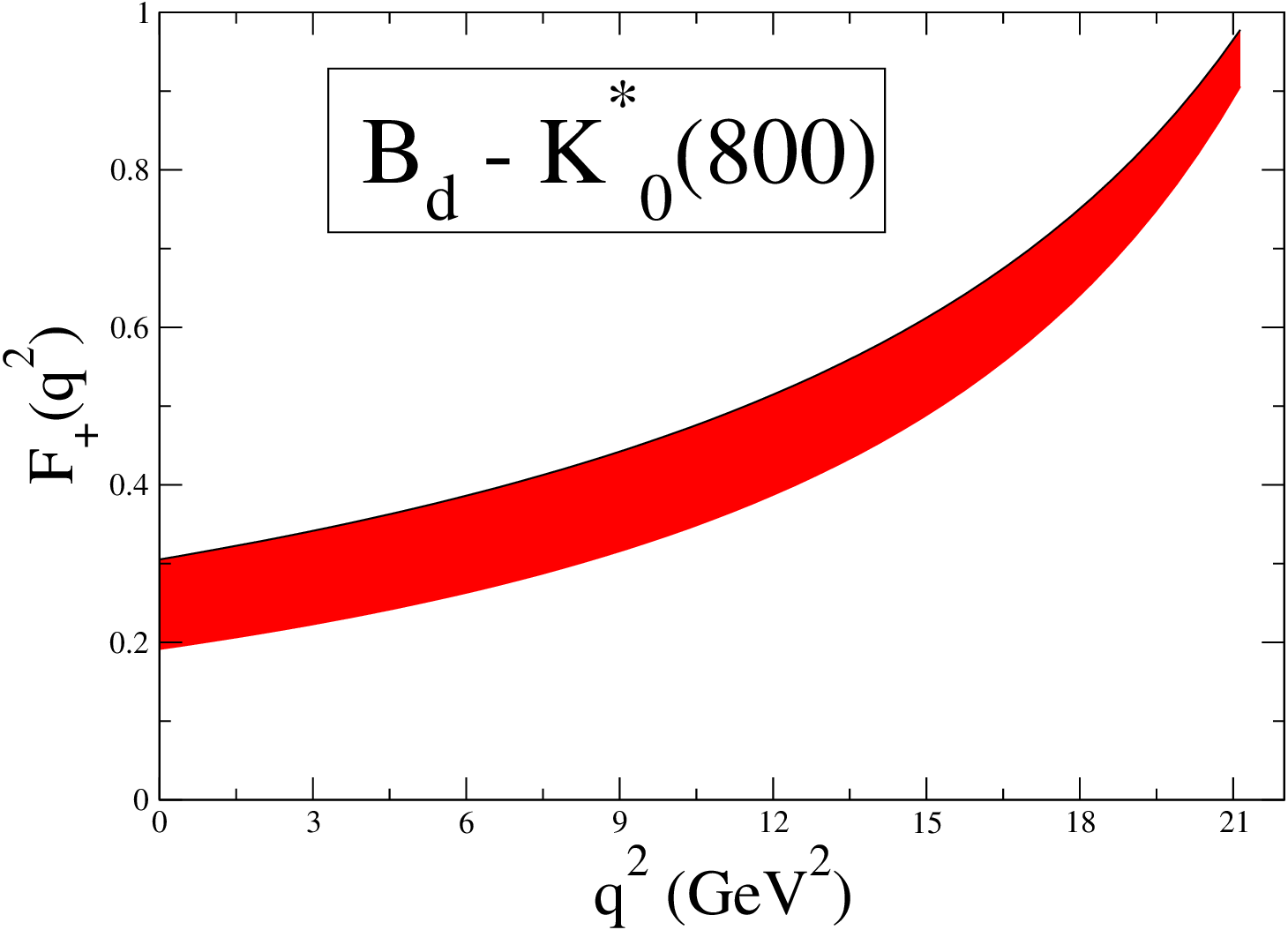}
\\[6ex]
\includegraphics[width=0.40\textwidth]{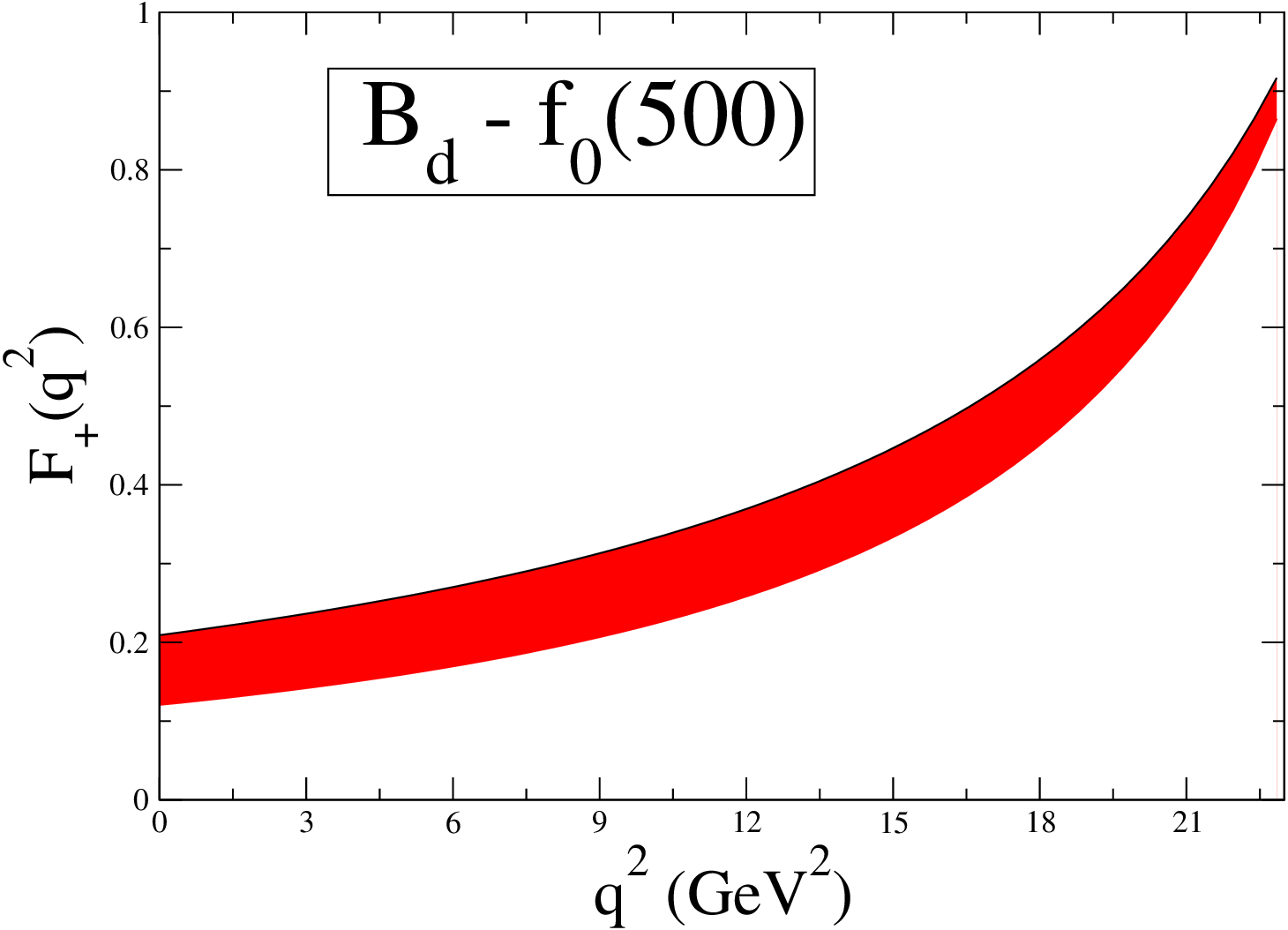}  \qquad & 
\end{tabular}
\end{center}
\caption{\label{fig:Fp}
The $F_+(q^2)$ form factors for the $b-u$, $b-d$, and $b-s$ transitions.
The upper and lower edges correspond to the values 
$\Lambda_S=1.5$~GeV and  $\Lambda_S=0.8$~GeV, respectively.  
}
\end{figure}

In Figs.~\ref{fig:Fp} and \ref{fig:FT}, we plot our calculated 
$F_{+}(q^{2})$ and $F_{T}(q^{2})$ form factors in the entire kinematical range 
$0 \leq q^{2} \leq q^{2}_{\rm max}$. Since the behavior of $-F_-(q^2)$
is very similar to that of $F_{+}(q^{2})$, we do not display them.
One can see that the form factors are more sensitive to the choice
of  $\Lambda_S$ at small $q^2$ and less so near zero recoil.

We are going to explore the influence of the intermediate scalar $K^\ast_0$
meson on the angular decay distribution of the cascade decay
$B\to K\pi+\mu^+\mu^-$. Therefore, we give the maximum values of the
form factors in  Table~\ref{tab:F(0)} and  the branching ratios in 
Table~\ref{tab:branching} obtained for $\Lambda_S=1.5$~GeV.
The results for the $e$ mode are almost identical to those 
of the $\mu$ mode and will not be shown separately.
Since the ratio $|V_{td}|/|V_{ts}|\approx 0.21$ is relatively small
we do not show the  branching ratios of the decays with
the $b-d$ transition.
We compare the obtained results with those from other approaches.
One can see that our values for the branching ratios are almost half of 
those from other approaches.    

\begin{figure}[ht]
\begin{center}
\includegraphics[width=0.70\textwidth]{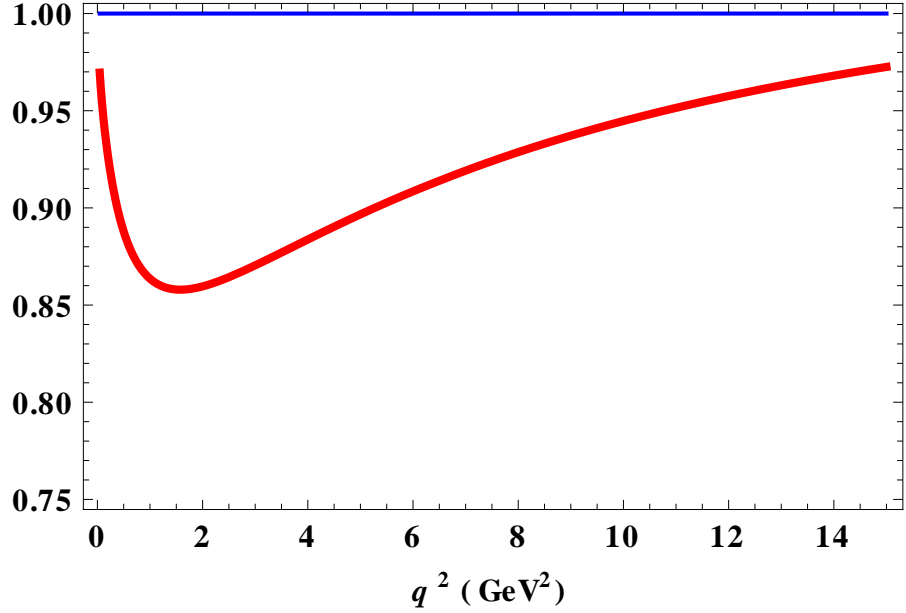}
\caption{The ratio of the differential decay rate 
$d\Gamma(B\to K^\ast(892)(\to K^0\pi^+)\mu^+\mu^-))$ 
to the full differential decay rate 
$ d\Gamma(B\to K^\ast(892)(\to K^+\pi^-)\mu^+\mu^-)
+ d\Gamma(B\to K^\ast_0(\to K^+\pi^-)\mu^+\mu^-)$.
}
\label{fig:ratio}
\end{center}
\end{figure}

Let us briefly discuss the impact of the scalar resonance $K^\ast_0$
on $B\to K^\ast(\to K\pi)\ell^+\ell^-$ decay. As is well known,
the narrow $K^\ast(892)$ vector resonance is described by 
a Breit-Wigner parametrization and the given cascade $B$ decay can be calculated
by using the narrow-width approximation. But this is not  true in the case
of the broad scalar $K^\ast_0(800)$ meson. There are several parametrizations
of the $K-\pi$ line shapes in the literature; see, for instance,
the discussion in Ref.~\cite{Doring:2013wka}. For the time  being we will use 
the parametrization accepted in Ref.~\cite{Meissner:2013pba}, 
the integrated value of which in the $K^\ast$-resonance region is equal to
\be
\int_{ (m_{K^\ast} - \delta_m)^2 }^{ (m_{K^\ast}+\delta_m)^2 } 
dm_{K\pi}^2
|L_S(m_{K\pi}^2)|^2 = 0.17, \quad\text{where}\quad \delta_m = 100\,\text{MeV}.
\label{eq:scale}
\en
Then, we scale the calculated value for the differential decay rate 
$d\Gamma(B\to K^\ast_0(800)\mu^+\mu^-)$ by this factor and compare it 
with that for $B\to K(892)\mu^+\mu^-)$ decay. We display the behavior
of the ratio
\be
R(q^2)=
\frac{2/3\, d\Gamma(B\to K^\ast(892)\mu^+\mu^-)}
{   2/3\, d\Gamma(B\to K^\ast(892)\mu^+\mu^-) 
 + 0.17 d\Gamma(B\to K^\ast_0(800)\mu^+\mu^-)}
\en
in Fig.\ref{fig:ratio}, which may be compared with the finding
of Ref.~\cite{Becirevic:2012dp}. The integrated ratio
(the numerator and denominator are integrated separately
in the full kinematical region of $q^2$ $\!$)  gives
a size for the $S$-wave pollution to the
branching ratio of the $B\to K^\ast\ell^+\ell^-$ decay
of about 6$\%$.

\begin{table}
\begin{center}
\def\arraystretch{1.5}
\begin{tabular}{|c|c|c|c|c|c|c|c|c|}
\hline
$B-S$  & $F(0)$  & This work &    \cite{Wang:2014vra}  &   \cite{Sun:2010nv} & 
\cite{Colangelo:1995jv} & \cite{Li:2008tk} & \cite{Ghahramany:2009zz} & 
\cite{ElBennich:2008xy}
\\    
\hline
$B^0_d-a^+_0(980)$ & $F_+(0)$ & 0.192 & 0.58 & 0.56 &&&&\\
%                  & $F_T(0)$ & 0.182 & 0.78 &      &&&&\\
\hline
$B^0_s-K^{\ast\,+}_0(800)$  & $F_+(0)$ & 0.274 & 0.44 & 0.53 &&&& \\
%                         & $F_T(0)$  & 0.292 & 0.60 &     &&&& \\
\hline
$B^0_s-f_0(980)$  & $F_+(0)$ & 0.254 & 0.45 & 0.44 & 0.19 & 0.35 & 0.12 & 0.40\\
                 & $F_T(0)$ & 0.285 & 0.60 & 0.58 & 0.23 & 0.40 &-0.08 &   \\
\hline
$B^0_d-K^{\ast\,0}_0(800)$  & $F_+(0)$ & 0.306 & 0.50 & 0.46 &&&&\\
                         & $F_T(0)$ & 0.306 & 0.67 & 0.58 &&&&\\
\hline
$B^0_d-f_0(500)$  & $F_+(0)$ & 0.210 & & &&&&\\
                 & $F_T(0)$ & 0.203 & & &&&& \\
\hline
\end{tabular}
\end{center}
\caption{ The values of the form factors at $q^2=0$ 
in the covariant quark model ($\Lambda_S=1.5$~GeV) and other
approaches. }
\label{tab:F(0)}
\end{table}

\begin{table}
\begin{center}
\def\arraystretch{1.5}
\begin{tabular}{|l|c|c|c|c|}\hline
Decay modes  &  \multicolumn{4}{c|}{Branching fractions} \\
\cline{2-5} 
& This work     & \cite{Wang:2014upa}
&  \cite{Colangelo:1995jv} & \cite{Li:2008tk} \\
& ($\Lambda_S=1.5$~GeV)  &&& \\
 \hline\hline
  $ B_d^0\to a^+_0(980) \mu^-\bar\nu_\mu $  & 
$0.52\times 10^{-4}$ & $ (2.74\pm 0.40) \times 10^{-4}$ && $1.84 \times 10^{-4}$ 
\\
  $ B_d^0\to a^+_{0}(980) \tau^-\bar\nu_\tau$ &
$0.11\times 10^{-4}$ & $ (1.31\pm 0.23) \times 10^{-4}$  && $1.01 \times 10^{-4}$ 
\\      
\hline
$B_s^0\to  K_0^{\ast\,+}(800) \mu^-\bar\nu_\mu $  &
$1.23\times 10^{-4}$   &  $(2.06\pm 0.31)\times 10^{-4}$ && $1.42 \times 10^{-4}$ 
\\
$B_s^0\to  K_0^{\ast\,+}(800) \tau^-\bar\nu_\tau$  &
$0.25\times 10^{-4}$  & $(1.07\pm 0.19)\times 10^{-4}$ && $0.88 \times 10^{-4}$
\\      
\hline\hline
$B_d^0\to  K_0^{\ast\,0}(800) \mu^+\mu^-$   & 
$3.47\times 10^{-7}$   &  $(7.31\pm 1.21)\times 10^{-7}$  &&
\\
$B_d^0\to  K_0^{\ast\,0}(800) \tau^+\tau^-$ & 
$0.61\times 10^{-7}$  & $(1.33\pm 0.36)\times 10^{-7}$  &&
\\      
\hline
$B^0_s\to f_{0}(980)\mu^{+}\mu^{-} $   & 
$2.45\times 10^{-7}$  & $(5.14\pm 0.78)\times 10^{-7}$    &  
$0.95\times 10^{-7}$ &$5.21\times 10^{-7}$ 
\\
$B^0_s\to f_{0}(980)\tau^{+}\tau^{-} $ & 
$0.42\times 10^{-7}$  & $(0.74\pm 0.17) \times 10^{-7}$ & 
$1.1\times 10^{-7}$ & $0.38\times 10^{-7}$  
\\
\hline\hline
$B^0_d\to K_0^{\ast\,0}(800) \bar\nu\nu$ & 
$2.53\times 10^{-6}$  & $(6.30\pm 0.97)\times 10^{-6}$ && 
\\
$B^0_s\to f_{0}(980)\bar\nu\nu $       & 
$1.79\times 10^{-6}$  & $ (4.39\pm 0.63)\times 10^{-6}$ & $0.87\times 10^{-6}$ & \\
\hline
\end{tabular}
\end{center}
\caption{ The branching fractions for the semileptonic and rare
$B$ decays into light scalar mesons and lepton pairs.}
\label{tab:branching}.
\end{table}

\section*{Acknowledgments}

We thank Pietro Santorelli for providing us with
the last fit of the parameters in the covariant quark model.
We would also like to thank Juergen K\"orner and
Valery Lyubovitskij for many useful discussions 
of $B$-physics facets related to the subject of this
paper. We are grateful to Wei Wang and David Straub
for pointing out the relevant references.

%\clearpage

\vspace*{1cm}

\vspace*{1cm}

\begin{figure}[ht]
\begin{center}
\hspace*{-0.5cm}
\begin{tabular}{lr}
%\includegraphics[width=0.40\textwidth]{Bda0980_FT.eps}  \qquad &\qquad
%\includegraphics[width=0.40\textwidth]{BsK0800_FT.eps} 
%\\[6ex]
\includegraphics[width=0.40\textwidth]{Bsf0980_FT.eps}  \qquad & \qquad
\includegraphics[width=0.40\textwidth]{BdK0800_FT.eps}
\\[6ex]
\includegraphics[width=0.40\textwidth]{Bdf0500_FT.eps}  \qquad & 
\end{tabular}
\end{center}
\caption{\label{fig:FT}
The $F_T(q^2)$ form factors for the $b-u$, $b-d$, and $b-s$ transitions.
The upper and lower edges correspond to the values 
$\Lambda_S=1.5$~GeV and  $\Lambda_S=0.8$~GeV, respectively.  
}
\end{figure}

\ed